\documentclass[preprint,12pt]{elsarticle}

\usepackage{amssymb}
\usepackage{amsmath}

\DeclareMathOperator*{\argmax}{arg\,max}
\usepackage{multirow}
\usepackage{listings} 
\usepackage{xcolor}
\usepackage{makecell}
\usepackage{booktabs}
\usepackage{soul}

\usepackage{float}
\usepackage[most]{tcolorbox}
\usepackage{adjustbox} 
\usepackage{hyperref}

\soulregister{\citep}{7}
\soulregister{\citeauthor}{7}
\soulregister{\footnote}{7}
\soulregister{\ref}{7}

\journal{Journal of Systems and Software}

\begin{document}

\begin{frontmatter}





\title{Variational Prefix Tuning for Diverse and Accurate Code Summarization Using Pre-trained Language Models} 


\author[1]{Junda Zhao} 


\affiliation[1,2,3]{organization={Department of Mechanical and Industrial Engineering, University of Toronto},
            addressline={5 King’s College Road}, 
            city={Toronto},
            postcode={M5S 3G8}, 
            state={Ontario},
            country={Canada}}

\author[2]{Yuliang Song} 

\author[3]{Eldan Cohen} 

\begin{abstract}
Recent advancements in source code summarization have leveraged transformer-based pre-trained models, including Large Language Models of Code (LLMCs), to automate and improve the generation of code summaries. However, existing methods often focus on generating a single high-quality summary for a given source code, neglecting scenarios where the generated summary might be inadequate and alternative options are needed. In this paper, we introduce Variational Prefix Tuning (VPT), a novel approach that enhances pre-trained models' ability to generate diverse yet accurate sets of summaries, allowing the user to choose the most suitable one for the given source code. Our method integrates a Conditional Variational Autoencoder (CVAE) framework as a modular component into pre-trained models, enabling us to model the distribution of observed target summaries and sample continuous embeddings to be used as prefixes to steer the generation of diverse outputs during decoding. Importantly, we construct our method in a parameter-efficient manner, eliminating the need for expensive model retraining, especially when using LLMCs. Furthermore, we employ a bi-criteria reranking method to select a subset of generated summaries, optimizing both the diversity and the accuracy of the options presented to users. We present extensive experimental evaluations using widely used datasets and current state-of-the-art pre-trained code summarization models to demonstrate the effectiveness of our approach and its adaptability across models.
\end{abstract}


\begin{highlights}
\item The first work that explores diverse and accurate code summarization using Large Language Models of Code (LLMCs).
\item We propose Variational Prefix Tuning (VPT), a novel approach that integrates Conditional Variational AutoEncoder (CVAE) as a modular component within LLMCs in a parameter-efficient manner, thus eliminating the need for costly retraining of the LLMC.
\item We demonstrate the adaptability of VPT by applying it to a variety of transformer-based pre-trained code summarization models.
\item Extensive experiments show VPT's capability to enhance LLMCs' ability to generate accurate summaries with improved diversity.
\end{highlights}


\end{frontmatter}



\section{Introduction}
\label{Introduction}

In the domain of software engineering, the task of code summarization is a crucial bridge between complex code and human comprehension \citep{MeasuringProgramCom}, providing a concise, human-interpretable explanation of the code's purpose. As software engineering practices evolve, the importance of code summarization has become increasingly recognized for improving code readability and maintainability \citep{Human_Study_com_code_sum}. Previous studies have shown that effective code summarization enables developers to quickly grasp the functionality of code without delving into detailed source code \citep{SummaryHelpsUnderstanding}. Conversely, the absence or inadequacy of code summaries can significantly impede development progress \citep{DeepCom}.

Recent work in the field of code summarization has predominantly leveraged neural networks as the primary tool for generating summaries from source code, yielding more robust models than traditional rule-based or information retrieval-based methods \citep{DeepCom}. Initial models relied on Recurrent Neural Networks (RNNs), such as Long Short-Term Memory (LSTM) networks, but the field has since shifted with the introduction of the Transformer model, which quickly emerged as the state-of-the-art \citep{NeuralCodeSum}. Transformers soon became the dominant architecture for code summarization tasks, with several powerful new models being developed based on this architecture \citep{SCRIPT, SG-transformer, wu-etal-sit-transformer}. Following the recent success of transformer-based Large Language Models (LLM), these models have been successfully applied to code-related tasks, including code summarization \citep{feng-etal-2020-codebert, wang2023codet5, rozière2024codeLLAMA}. Large Language Models of Code (LLMC) are generally pre-trained on extensive datasets comprising both code and natural language and can be further fine-tuned to fulfill specific task requirements.

Typically, pre-trained models for code summarization aim to generate a single summary for a given input source code, with their performance evaluated based on the accuracy of this single generated summary. This observation is grounded in our comprehensive literature review, which examined over 100 publications and was anchored in the latest survey on automatic source code summarization \citep{ReviewAutoCodeSumm}, covering works from the past ten years up to June 2024. After thoroughly reviewing all of the code summarization literature referenced in the survey, we found that: (1) none of the studies have evaluated the ability to generate a set of multiple summaries such that at least one of them is adequate for the user, and (2) no prior study has considered the task of generating a diverse set of summaries. However, consider a scenario where a developer seeks to generate a summary for a complex piece of code automatically. If the model produces an inadequate summary, the developer has no other choice but to reject it. To address this problem, we propose generating a set of diverse summaries that is more likely to contain an adequate summary for the user.

Existing methods for generating multiple outputs from a pre-trained model face challenges in producing a set of diverse yet accurate summaries. Techniques like beam search often result in outputs with only minor differences \citep{macherey-etal-2008-lattice, tromble-etal-2008-lattice, kumar-byrne-2004-minimum}, while methods emphasizing diversity, such as sampling and diverse beam search, tend to suffer from a significant loss in accuracy, as we later demonstrate in our experimental results.

Variational Autoencoders (VAE) \citep{VAE} and Conditional Variational Autoencoders (CVAE) \citep{CVAE} have been successful in introducing diversity into neural network models in a variety of generation tasks \citep{wang2017diverseandaccurate, lin2020variationalTransformer, CreativityDiverseQuestion}. However, to the best of our knowledge, this paradigm has not been applied to the code summarization task. While recent work has considered integrating VAE/CVAE with transformers \citep{T-CVAE, fang2021transformerbasedCVAE, TransformerVAEmusiclearning}, these approaches often require a complete restructuring of the transformer model, necessitating retraining from scratch. Given that pre-trained models today typically contain a significant number of parameters and require extensive resources to train, retraining the entire model may not be practical due to limited computational resources.

With this in mind, we draw inspiration from Prefix Tuning, a \emph{parameter-efficient fine-tuning method} (PEFT) that employs a separate encoder to learn a set of vectors as task-specific prefixes. These prefixes guide the generation process while keeping the rest of the pre-trained model frozen during training, thus requiring significantly less computational resources. 

\begin{figure*}[ht]
    \centering
    \includegraphics[width=1.\linewidth]{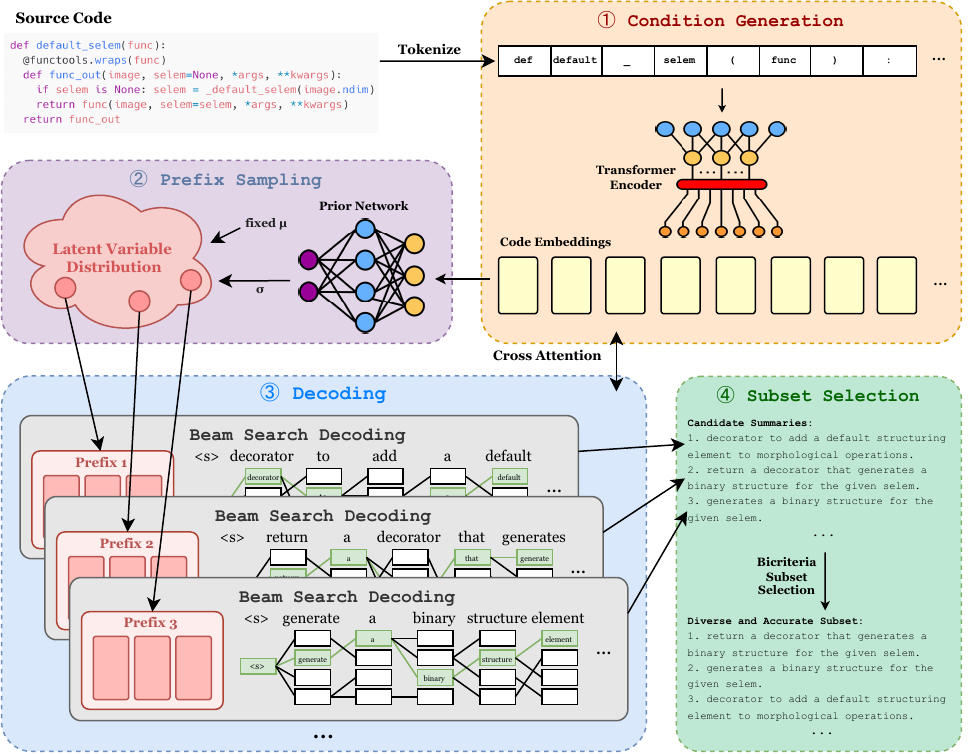}
    \caption{An overview for using VPT to generate a diverse and accurate set of code summaries.}
    \label{fig:overview}
\end{figure*}

Using a similar integration paradigm of prefix tuning, we introduce our novel approach, Variational Prefix Tuning (VPT), which integrates the CVAE framework into pre-trained models in a parameter-efficient way. Figure~\ref{fig:overview} provides an overview of our method. Instead of learning a set of fixed task-specific prefixes, we use a set of stochastic continuous prefixes generated by the CVAE framework to steer the generation. Each different set of continuous prefixes, when used during the decoding process, can result in varying outputs given the same input source code. By incorporating the CVAE framework, our method enhances the capabilities of pre-trained models to generate a diverse set of outputs that are more likely to include an accurate summary. To further improve both diversity and accuracy of the returned set of summaries, we employ a two-step process: VPT first generates a larger pool of candidate outputs, from which we apply bi-criteria optimization to select a subset of desired size that maximizes quality and accuracy.

In this paper, we detail the architecture of the proposed approach and provide comprehensive experiments demonstrating the approach's ability to empower pre-trained models to produce diverse yet accurate source code summaries. In summary, we make the following contributions:

\begin{enumerate}
\item We propose \emph{Variational Prefix Tuning}, a novel parameter-efficient approach that integrates a CVAE framework as a modular component within pre-trained transformer-based models, enhancing their ability to generate diverse and accurate summaries without the need for costly retraining. Further, we enhance our approach with a bi-criteria subset selection method to maximize both summary quality and diversity in the selected set of summaries. 
\item To demonstrate the adaptability of our method across various model architectures and scales, we integrate our model into a variety of transformer-based pre-trained code summarization models, including state-of-the-art LLMCs.
\item We conduct extensive experiments to demonstrate VPT's ability to generate accurate summaries with enhanced diversity. In particular, we demonstrate that it outperforms various strategies for generating multiple outputs, including ones designed for diverse generation. Our replication package is available at \url{https://github.com/jundaz/VPT}.
\end{enumerate}

\section{Preliminaries}
\subsection{Problem Formulation}
Source code summarization in our model is approached as a sequence-to-sequence (seq2seq) problem, where both the source code input and the natural language summary output are considered as sequences of tokenized elements. The primary goal of our model is to translate a sequence of source code tokens, represented as $\mathbf{x}=(x_1,\ldots,x_n)$, into a coherent and concise natural language summary, represented as a sequence of natural language tokens $\mathbf{y}=(y_1,\ldots,y_m)$. This summary should accurately reflect the core functionality and intent of the source code.

The problem we address specifically is the generation of a diverse yet accurate set of summaries from the input source code. We formalize this problem similarly to diverse generation tasks in other domains, such as image captioning, question generation and conversation generation \citep{vijayakumar2018diverse, wang2017diverseandaccurate, aneja2019sequentiallatentdiversecaption, CreativityDiverseQuestion, diverseconversation}. Our objective is to maximize the likelihood that the generated set of summaries contains at least one correct description that accurately reflects the code's functionality. To achieve this, our approach generates summaries that are both semantically and lexically diverse, thereby covering a broader spectrum of possible accurate summaries for the given source code. We evaluate our model based on its ability to generate such a diverse and accurate set of summaries.

\subsection{Large Language Model of Code (LLMC)}
Current mainstream Large Language Models for Code (LLMC) predominantly utilize Multi-head Attention Mechanisms as proposed by \citeauthor{transformer} \cite{transformer}. These transformer-based models typically feature several layers of Multi-head Attention, equipped with a substantial number of trainable parameters. This architecture enables the models to effectively capture complex term dependencies and contextual information within both source code and natural language. The multi-head attention process is mathematically represented as follows:
\begin{align}
\text{Attention}(Q,K,V) &= \text{softmax}\left(\frac{QK^T}{\sqrt{d_k}} \right)V\\
\text{MultiHead}(Q,K,V) &= \text{Concat}(head_1, \dots, head_h)W^O\\
\text{where} \ head_i &= \text{Attention}(QW^Q_i, KW^K_i, VW^V_i)
\end{align}

Transformer-based large pre-trained models for code can be divided into three main categories based on their architectures:

\begin{enumerate}
\item \textbf{Encoder-Only}: BERT-like models utilize Bidirectional Multi-head Attention. Their primary pre-training tasks include Masked Language Modeling (MLM) and other tasks specifically designed for code \citep{feng-etal-2020-codebert}. Notable models include CodeBERT \citep{feng-etal-2020-codebert}, GraphCodeBERT \citep{guo2021graphcodebert}, and ContraBERT \citep{liu2023contrabert}. Encoder-only models are often used for classification or retrieval tasks and can only perform generation tasks such as code summarization when attached to a decoder and fine-tuned.

\item \textbf{Decoder-Only}: GPT-like models employ Multi-head Attention with causal masks, preventing tokens from attending to future positions. They are mainly pre-trained in an autoregressive manner. Notable models include CodeGPT \citep{lu2021codegpt}, Codex \citep{chen2021codex}, and CodeLlama \citep{rozière2024codeLLAMA}. Decoder-only models can generally scale well and are often used for general-purpose generation tasks with prompting, as their large scale of parameters allows them to encode a wide range of knowledge.

\item \textbf{Encoder-Decoder}: These models maintain a transformer-like structure and treat input and output as seq2seq tasks, allowing for flexible pre-training tasks to be applied to both the encoder and the decoder. Notable models include CodeT5 \citep{wang2023codet5} and PLBART \citep{ahmad2021PLBART}. Encoder-decoder models have demonstrated their effectiveness on seq2seq tasks such as code summarization and translation \citep{wang2023codet5, transcoder}.
\end{enumerate}

\paragraph{CodeT5+}
\label{section:codet5+}
To demonstrate the effectiveness of the VPT method, we used CodeT5+, an encoder-decoder structured LLMC that has achieved state-of-the-art performance in code summarization \citep{wang2023codet5} as our primary model. Derived from the T5 model \citep{raffel2023exploring}, CodeT5+ is a scaled-up version of the original Transformer architecture. When applied to the task of source code summarization, the encoder processes code tokens with multi-head self-attention to capture complex relationships and encode them into contextual embeddings, while the decoder generates summaries by focusing on relevant parts in these contextual embeddings through cross-attention layers. While maintaining the encoder-decoder structure characteristic of Transformer models, it incorporates significantly more trainable parameters, enhancing its capability to understand both code and natural language.

CodeT5+ introduces a range of pre-training tasks specifically designed for coding contexts to strengthen its ability to process code-related inputs. These tasks include both uni-modal (code-only) and bi-modal (code + text) objectives. The uni-model tasks focus on masked language modeling (MLM) for the encoder, where the model predicts masked tokens in the input, and causal language modeling (CLM) for the decoder, where the model autoregressively predicts the next token in a sequence. For bi-modal tasks, Text-Code Contrastive Learning and Text-Code Matching are employed to align the representation space, pushing apart irrelevant code-text pairs and bringing semantically similar ones closer. Additionally, a sequence-to-sequence generation task is introduced, where the encoder processes input (code or text), and the decoder generates the corresponding output (text or code). Special prefix tokens, such as [Tdec] for Code-to-Text and [Cdec] for Text-to-Code, guide the decoder’s generation process \citep{wang2023codet5}. These tasks enhance CodeT5+'s ability to understand source code and generate accurate summaries.

\subsection{Variational Autoencoder (VAE)}

A Variational Autoencoder (VAE) \citep{VAE} is a generative model that consists of an encoder and a decoder, similar to a standard Autoencoder (AE). The encoder, denoted as $q_{\phi}(z|y)$, maps data $y$ to a latent space, and the decoder, $p_{\theta}(y|z)$, attempts to reconstruct the data $y$ from these latent variables. Unlike AEs, VAEs introduce a prior distribution $p(z)$ over the latent space and learn an approximate posterior distribution $q_{\phi}(z|y)$, enhancing their ability to model semantic features \citep{bowman-etal-2016-generating} and generate new and diverse data similar to observed samples. The optimization objective of VAE can be formulated as maximizing the Evidence Lower Bound (ELBO):
\begin{equation}
\argmax_{\phi, \theta}\{\mathbb{E}_{q_{\phi}(z|y)}[\log p_{\theta}(y|z)] - D_{KL}(q_{\phi}(z|y)||p(z))\}
\end{equation}
ELBO comprises a reconstruction term that encourages accurate input reconstruction and a Kullback-Leibler (KL) divergence term, a regularizer to align the learned posterior with the prior.

The Conditional VAE (CVAE) \citep{CVAE} extends VAEs into a conditional generative model containing 3 types of variables: input variable $x$, output variables $y$, and latent variables $z$ \citep{CVAE}. Both encoder and decoder of CVAE models are conditioned on input variable $x$, represented as $q_{\phi}(z|y, x)$ and $p_{\theta}(y|z, x)$, respectively. This setting allows for a diverse generation of outputs conditioned on the input context $x$. The adjusted ELBO can be formulated as follows:
\begin{equation}
\argmax_{\phi, \theta}\{\mathbb{E}_
{q_{\phi}(z|x, y)}[\log p_{\theta}(y|z, x)] - D_{KL}(q_{\phi}(z|y, x)||p(z|x))\}
\end{equation}

\subsection{Prefix-tuning}
Prefix-tuning \citep{li2021prefixtuning} is a \emph{parameter-efficient fine-tuning method} for LLMs that reduces computational resource requirements by freezing the model's parameters and learning a set of continuous vectors known as a prefix. Inspired by prompting and in-context learning, this technique uses the learned prefix to guide the decoder's generation process. This enables the LLM to produce outputs that meet specific task requirements without modifying all parameters, significantly reducing computational resource requirements compared to fine-tuning the entire model.

The primary mechanism includes a prefix encoder with two layers: an embedding layer that converts virtual tokens into continuous vectors, and a Multi-Layer Perceptron (MLP) layer that further processes these vectors. These transformed embeddings function as the prepended key and value pairs for the self-attention layer in the decoder, thus guiding the generation decisions. The optimization process aims to maximize the likelihood of generating the correct target token at each timestep $t$ given the set of continuous prefixes denoted as $y'$, which can be formulated as:
\begin{equation}
\argmax_{\mathbf{y'}} P(y_t | y', y_{0:t-1}, \mathbf{x})
\end{equation}

\section{Variational Prefix Tuning (VPT)}

We introduce \emph{Variational Prefix Tuning}, a novel method designed to enable pre-trained transformer-based code summarization models to generate diverse and accurate code summaries by incorporating a CVAE framework in a parameter-efficient manner. Similar to prefix tuning, we use a set of continuous prefixes to efficiently adapt a pre-trained model to a new setting. However, in contrast to prefix tuning, we do not rely on a set of fixed vectors. Instead, we incorporate a variational prefix encoder to model the distribution of target summaries within a regulated latent space. Sampled latent variables from this distribution serve as our variational prefixes, supporting a more diverse yet accurate generation. In the following, we discuss how the core components are integrated and work together to realize our objective of producing diverse and accurate code summaries. 
\begin{figure*}[ht]
    \centering
    \includegraphics[width=1.\linewidth]{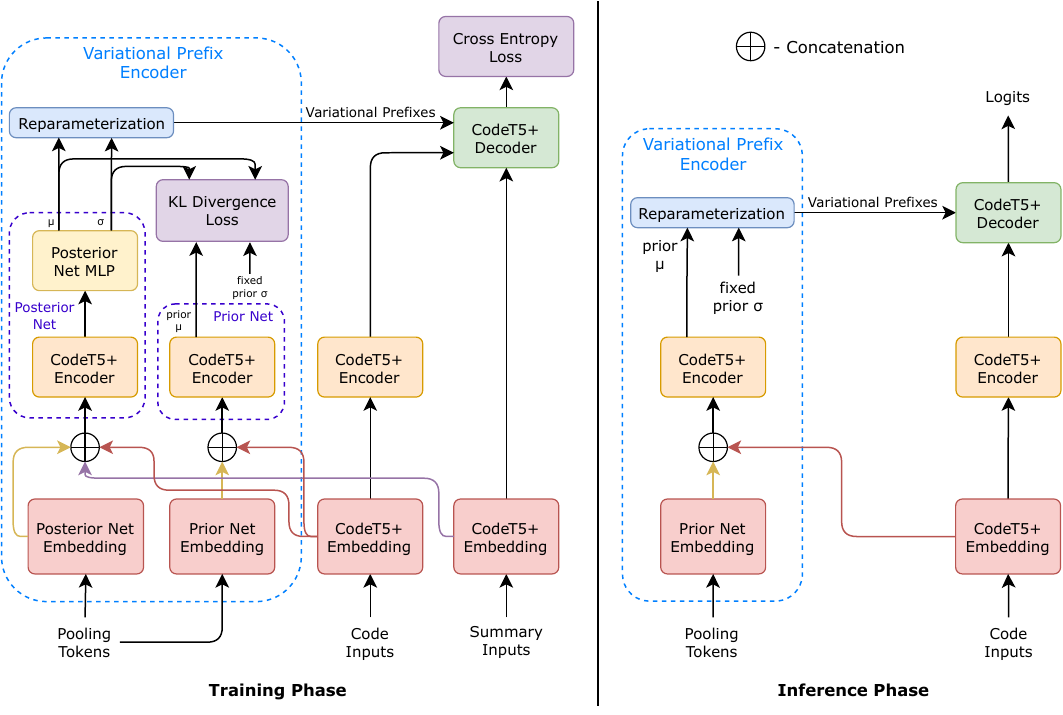}
    \caption{Detailed Model Architecture of VPT. }
    \label{fig:network structure}
\end{figure*}

\subsection{Backbone Model}  
For our approach, we adopted pre-trained transformer-based encoder-decoder models commonly used for code summarization tasks \citep{wang2023codet5, NeuralCodeSum, ahmad2021PLBART, SCRIPT, AST-trans} as the backbone of our method. The backbone model is responsible for processing source code tokens \(\mathbf{x}\) into contextual embeddings used as the conditional input and decodes the summaries out of the contextual embeddings together with the variational prefixes. As noted in Section~\ref{section:codet5+}, we selected CodeT5+ as our primary backbone model. Leveraging extensive pre-training data, the semantically rich contextual embeddings of the source code generated by the CodeT5+ encoder also allow us to parameterize a more informative prior distribution, as we discuss in the next section. 

\subsection{Prior Distribution}
\label{subsec:Prior}
The prior distribution, denoted as \( p_{\theta}(z|x) \), is a Gaussian distribution conditioned on the input source code, parameterized by a mean and a fixed standard deviation. This distribution regulates the posterior distribution during training and is used to sample latent variables during inference to steer generation. 

For our approach, we condition our prior distribution on the contextual embedding of the source code. Specifically, we utilize the frozen CodeT5+ encoder to generate embeddings that capture the semantic information of the source code, referred to as the "Prior Net" in Figure~\ref{fig:network structure}. These embeddings serve as the mean of the prior distribution. To achieve this, we use a set of trainable pooling tokens that aggregate information from the source code sequence, functioning similarly to the \textless CLS\textgreater\ token in BERT \citep{devlin2019bert}. The contextual embeddings of these pooling tokens, processed by the CodeT5+ encoder, parameterize the mean of the prior distribution.

Parameterizing the prior distribution in this manner is beneficial because the CodeT5+ encoder, with its multi-head self-attention layers, effectively encodes the source code by capturing contextual information. Having been pre-trained on a variety of code-related tasks, the encoder learns a code embedding space that provides a more informative and relevant prior for guiding the posterior distribution compared to a standard normal distribution with a fixed mean of 0. During training, the prior distribution’s standard deviation is fixed at 1, while during inference, it can be scaled to adjust the randomness of the generation process.

\subsection{Posterior distribution}
\label{sec:posterior net section}

The posterior distribution captures the distributional information of all summaries in the training set while being regulated by approximating the prior distribution during training. Consequently, it should be conditioned on both the source code \(\mathbf{x}\) and the summary \(\mathbf{y}\), and is denoted as \( q_{\phi}(z|y, x) \).

Similar to the prior distribution, we utilize the frozen encoder of the CodeT5+ model to generate contextual embeddings that parameterize the posterior distribution, referred to as the "Posterior Net" in Figure~\ref{fig:network structure}. We concatenate the source code and its corresponding summary into a single text sequence, using a separate set of trainable pooling tokens to aggregate information from the sequence. These tokens are processed by the CodeT5+ encoder, and the resulting embeddings are transformed into the mean and standard deviation of the posterior distribution via two MLP layers. Unlike the prior distribution, the standard deviation of the posterior is learned rather than fixed, which helps alleviate posterior collapse \citep{dontblameelbo}.

It is important to note that utilizing the frozen encoder of our backbone model, CodeT5+, to generate the posterior distribution is feasible due to its pre-training with bi-modal data. When applying our VPT method to other baseline models that are trained solely on the seq2seq task of generating natural language from source code, or with only uni-model data, we instead use a trainable multi-head self-attention module.

\subsection{Training and Inference}
We present the detailed model architecture and dataflow during both training and inference in Figure~\ref{fig:network structure}. 
During training, our optimization objective is:
\begin{equation}
    \argmax_{\phi, \theta}\{\mathbb{E}_{q_{\phi}(z|\mathbf{y}, \mathbf{x})}[\log p_{\theta}(\mathbf{y}|z, \mathbf{x})] - D_{KL}(q_{\phi}(z|\mathbf{y}, \mathbf{x})||p_{\theta}(z|\mathbf{x}))\}.
\end{equation}
The KL-divergence term, \( D_{KL}(q_{\phi}(z|\mathbf{y}, \mathbf{x})||p_{\theta}(z|\mathbf{x})) \), is computed between the prior distribution \( p_{\theta}(z|\mathbf{x}) \) and the posterior distribution \( q_{\phi}(z|\mathbf{y}, \mathbf{x}) \). To obtain these distributions, we use two sets of trainable pooling token embeddings, each processed by separate token embedding layers. These embeddings are then concatenated with the code embeddings (for the prior) and the code+summary embeddings (for the posterior) before being passed through their respective networks, generating the contextual embeddings that parameterize both distributions.

The reconstruction term, \( \mathbb{E}_{q_{\phi}(z|\mathbf{y}, \mathbf{x})}[\log p_{\theta}(\mathbf{y}|z, \mathbf{x})] \), is optimized by minimizing the cross-entropy loss for the next token prediction task, similar to training a standard transformer-based model with one key difference: we sample our latent \(z\) from the posterior distribution and use them as prefixes to guide summary sequence generation during decoding.

Posterior collapse is a common issue that happens during the training phase of VAE/CVAE models, where the signal from the posterior approximation becomes too weak, leading the decoder to ignore the latent variable \(z\) and thereby limiting generation diversity \citep{goyal2017zforcingtrainingstochasticrecurrent}. This issue is particularly pronounced with strong autoregressive decoders \citep{zhu-etal-2020-batch-norm-vae}. To address this, we employ two methods: \emph{Cyclic KL Weight}, which dynamically cycles the weight of the KL divergence loss between 0 to 1 during training \citep{cyclicKlAnnealing}; and \emph{VAE BatchNorm}, which applies batch normalization to the VAE layers that generate the mean \(\mu\) and standard deviation \(\sigma\) of the posterior distribution. \citep{ioffe2015batchnormalizationacceleratingdeep, kexuefm-7381}.

During inference, latent variables \(z\) are sampled from the prior distribution rather than the posterior distribution. These sampled latent variables serve as prefixes to guide summary generation.

\subsection{Beam Search}
To further boost the quality of each sampled prefix's generated summary. We apply beam search to each prefix's decoding process.

As we formulated in the previous part, the next token's probability in the decoded sequence can be modeled as conditional probability: $P(y_t | y_{0:t-1}, \mathbf{x})$ where \( y_t \) represents the token at position \( t \), \( y_{0:t-1} \) denotes the preceding tokens in the sequence, and \( \mathbf{x} \) is the input.

While finding the optimal sequence through exhaustive search is impractical due to computational constraints, greedy search that chooses the most likely token at each step usually leads to suboptimal global sequences. Beam search offers a more balanced solution by expanding and retaining the top $B$ (beam width) partial sequences at each step.

Let \( Y_{t-1} \) be the set of \( B \) best sequences at step \( t-1 \). Beam search then explores all possible extensions of these sequences by one token and retains the top \( B \) sequences based on their cumulative score. Formally, this update process can be formulated as:
\begin{equation}
\begin{split}
Y_t = \argmax_{y_{0,t},...,y_{B-1,t} \in Y_t} \bigg\{ \sum_{b=0}^{B-1} \log P(y_{b,t} | \mathbf{x}) \bigg\} \\
\text{s.t. } y_i \neq y_j \text{ for all } i, j \text{ with } i \neq j \text{ and } i, j \in \{0, ..., B-1\}
\end{split}
\end{equation}
This method efficiently searches the space of possible sequences, ultimately returning the beam with the highest log probability as the approximate most likely sequence.

\subsection{Bi-Criteria Subset Selection}
\label{Bi-Criteria Subset Selection}

We adopt a two-stage procedure to enhance further the diversity and accuracy of the final set of output summaries presented to the users. In the first stage, we generate a larger initial set of summaries, \( S \), of size \( N \). In the second stage, we apply a bi-criteria subset selection method, following the methodology in \citeauthor{zhong2024bi-criteria} \citep{zhong2024bi-criteria}, to select a subset \( s \) of size \( U \), where \( U < N \). This selection balances both accuracy and diversity by optimizing the following objective:
\begin{equation}
\argmax_{s}\{\alpha g(s) + \beta h(s)\},  s\in S, |s| = U, |S| = N
\end{equation}
where $g(s)$ represents the quality measurement and $h(s)$ represents the diversity measurement. The parameters $\alpha$ and $\beta$ control the trade-off between diversity and quality. In our use case, the quality $g(s)$ is measured by the sum of all length $T$ normalized log probability of the summaries within the set $s$, predicted by the backbone model. This is denoted as:
\begin{align}
    g(s) = \sum_{\mathbf{y}\in s} \frac{\sum_{t=1}^{T} \log P(y_{t} | y_{0:t-1}, \mathbf{x})}{T}
\end{align}
The diversity is measured by the sum of all pairwise distances, computed using one minus the BLEU-4 score \citep{bleu} between each pair of summaries within set $s$, denoted as:
\begin{align}
    h(s) = \sum_{\mathbf{y}_i, \mathbf{y}_j\in s} 1 - \text{BLEU}(\mathbf{y}_i, \mathbf{y}_j)
\end{align}
This subset selection procedure further enhances the likelihood of containing a correct summary within the generated set of summaries.

\subsection{Computational Efficiency of VPT}
\begin{table}[ht]

    \centering
    \begin{tabular}{r|r|r}
    \hline
     & Total Parameters & Trainable Parameters \\
    \hline
    Full finetuning & 222882048 & 222882048 \\ 
    VPT             & 247088640 & 24206592 \\
    \hline
    \end{tabular}
    \caption{Comparison of total and trainable parameters for full fine-tuning and VPT.}
    \label{tab:parameters}

\end{table}

In this section, we discuss the computational efficiency of VPT during both training and inference. Regarding training efficiency, as VPT is applied to a fully fine-tuned model, we note that the additional computational cost is very limited due to the significantly reduced number of trainable parameters. As shown in Table~\ref{tab:parameters}, the number of trainable parameters in VPT constitutes approximately 10.8\% of those required for full fine-tuning, demonstrating that our approach is computationally efficient to implement. Regarding inference efficiency, VPT adds a fixed, small number of extra tokens (only two in our case). Similar to standard prefix tuning, this overhead is negligible because the attention computation over the entire prefix is efficiently parallelized on GPUs \citep{li2021prefixtuning}.

\section{Experiment Setup}

\subsection{Research Questions}

We investigate the effectiveness of VPT for diverse code summarization through the following research questions:

\begin{enumerate}
    \item \textbf{RQ1: Does VPT increase the probability that a generated set of summaries includes a superior summary compared to baseline decoding methods?}\\
    We evaluate whether VPT produces sets of code summaries that are more likely to include an accurate summary than those generated by the baseline methods, using a comprehensive suite of metrics.
    
    \item \textbf{RQ2: Are the code summaries produced by VPT more diverse than those generated by baseline decoding methods?}\\
    We evaluate VPT's capability to generate more diverse sets of code summaries compared to baseline methods, employing a suite of diversity metrics.
    
    \item \textbf{RQ3: How effectively does VPT perform when applied to various pre-trained models for code summarization, and how does it compare with other PEFT methods and SOTA general-purpose LLM?}\\
    We evaluate the effectiveness and adaptability of VPT by integrating it into multiple pre-trained code summarization models. Furthermore, our experiments compare VPT to both CodeLlama, a SOTA decoder-only LLMC, finetuned for code summarization using LoRA, and a general-purpose LLM GPT-4o with few-shot learning.
    
    \item \textbf{RQ4: What is the impact of each additional component integrated into VPT (i.e., beam search, prior network, and bi-criteria subset selection) on its overall performance?}\\
    We conduct an ablation study, examining the individual contributions of beam search, the prior network, and the bi-criteria subset selection to VPT's overall performance.
\end{enumerate}





\subsection{Datasets}
Our analysis is based on two distinct datasets: the Java dataset from \citeauthor{transferAPI} \citep{transferAPI}, and the Python dataset, originally compiled by \citeauthor{pythonDataBasic} \citep{pythonDataBasic} and further refined by \citeauthor{rlhybrid} \citep{rlhybrid}. Table~\ref{tab:dataset-stats} outlines the specifics of the employed dataset. These two datasets are widely used as benchmark datasets for code summarization \citep{NeuralCodeSum, SCRIPT, SG-transformer, AST-trans, sun2023extractiveandabstractive}.

\begin{table}[htbp]
\centering
\caption{Dataset Statistics for Java and Python}
\label{tab:dataset-stats}
\begin{tabular}{l|c|c}
\hline
\textbf{Statistic} & \textbf{Java} & \textbf{Python} \\
\hline
Train instances & 69,708 & 55,538 \\
Validation instances & 8,714 & 18,505 \\
Test instances & 8,714 & 18,502 \\
\hline
\end{tabular}
\end{table}

\subsection{Evaluation Metrics}
\subsubsection{Accuracy Metrics}

To evaluate our model's performance in generating code summaries, we focused on comparing the generated summaries with the ground-truth summaries, as well as on assessing the semantic similarity between the generated summaries and the input code. For evaluating the accuracy of generated summaries against the reference texts, we employ three widely-used evaluation metrics for assessing the quality of generated summaries based on reference summaries, namely BLEU-4 \citep{bleu}, ROUGE-L \citep{rouge}, and METEOR \citep{meteor}. In addition, we incorporate BERTScore \citep{bertscore}, a more recent metric that utilizes BERT-based pre-trained language models to evaluate the semantic similarity between a predicted summary and its reference. To evaluate the semantic similarity between the generated summaries and the input code directly, we adopted the SIDE metrics \citep{SIDE_Score}. Similar to BERTScore, SIDE uses a BERT-based pre-trained language model to measure semantic similarity. We follow the original fine-tuning pipeline for SIDE, fine-tuning the scoring model on our dataset to accurately assess the semantic correspondence between the input code and the generated summary. Specifically, we utilize the official training script provided by the authors of SIDE\footnote{https://github.com/antonio-mastropaolo/code-summarization-metric/blob/main/train-hard-negatives.py} to fine-tune two separate MPNet models—one for the Java dataset and another for Python—using their contrastive learning pipeline. This diverse set of metrics ensures a comprehensive evaluation of our model's summarization performance.

Consistent with previous work on diverse text generation, we focus on \emph{Oracle} metrics, the established methods for evaluating diverse generation \citep{vijayakumar2018diverse, kool2019stochastic, dualcontrastive, aneja2019sequentiallatentdiversecaption, mao2015deepcaption}. The goal of diverse generation is to create a list of diverse options that are more likely to contain an accurate output, measured by its closeness to the ground-truth summary. Aligned with this premise, Oracle metrics evaluate a set of candidates by selecting the most accurate one according to each metric. This approach assesses how likely the diverse generation method is to generate an accurate summary within the set of distinct outputs, and the evaluation results represent an upper bound on the performance of any automated re-ranking technique \citep{wang2017diverseandaccurate}.

\subsubsection{Diversity Metrics}

In addition to evaluating accuracy metrics, we assess the diversity of our generated summaries using metrics commonly adopted in prior studies on diverse generation \citep{diverseconversation, wang2017diverseandaccurate, CreativityDiverseQuestion}.

Our first diversity metric is the distinct n-gram ratio \citep{DistinctN-gram}, where an n-gram is defined as a contiguous sequence of n tokens (typically words) in the text. Formally, given a set of generated summaries $S = \{ s_1, s_2, \dots, s_k \}$ for a particular input, let $U_n(S)$ denote the set of unique $n$-grams across all summaries in $S$, and let $T_n(S) = \sum_{i=1}^{k} T_n(s_i)$ be the total number of $n$-grams across the entire set. Here, $T_n(s_i)$ is the count of $n$-grams in the individual summary $s_i$. The distinct-$n$ score is then defined as:

\begin{equation}
    \text{D-}n = \frac{|U_n(S)|}{T_n(S)}
\end{equation}

For $n = 1$ and $n = 2$, this formulation yields the distinct unigram (D-1) and distinct bigram (D-2) ratios, respectively. For each input source code, we calculate D-1 and D-2 for every generated summary and then average these values to obtain the diversity measure for this set of summaries. A higher Distinct N-gram ratio indicates a higher diversity in generation.

The other diversity metric we use is the Self-BLEU metric \citep{zhu2018texygen}. The Self-BLEU metric compares each summary against all others by calculating the pairwise BLEU score of each summary within the set of generated summaries and then averaging these BLEU scores to determine the overall diversity for the generated set. Specifically, for a given input source code, let $ S = \{ s_1, s_2, \dots, s_k \} $be the set of generated summaries. The Self-BLEU for the generated set is defined as:

\begin{equation}
    \text{Self-BLEU}(S) = \frac{1}{k(k-1)} \sum_{i=1}^{k} \sum_{\substack{j=1 \\ j \neq i}}^{k} BLEU(s_i, \{ s_j \})
\end{equation}

Based on this definition, a lower Self-BLEU score indicates greater diversity among the generated summaries.

By adopting these two metrics together, we demonstrate that our approach, the VPT, is capable of producing a diverse set of code summaries.

\subsection{Baselines}

\subsubsection{Diverse Decoding Methods}
We also consider the following diverse generation methods, which have proven effective in generating sets of diverse and accurate outputs in other tasks.

\begin{enumerate}
\item \textbf{Beam Search}: Instead of using beam search to return only the most probable sequence to improve the quality of a single output, as mentioned earlier, beam search can also be used to generate a set of distinct summaries. This is achieved by returning the sequences from multiple beams, each representing a different plausible summary.
\item \textbf{Sampling}: Sampling from output logits as a probability distribution is commonly used in LLMs to generate diverse outputs. By adjusting the temperature parameter $T$, this method controls the randomness of token selection, with higher temperatures increasing diversity and lower temperatures leading to more deterministic outputs. For our experiment, we fine-tune temperature $T$ based on performance on a held-out validation set. 
\item \textbf{Stochastic Beam Search} \citep{kool2019stochastic}: Stochastic Beam Search (SBS) enhances traditional beam search by introducing randomness into the selection process to increase diversity. This method perturbs log-probabilities with random variations, balancing randomness and determinism. The level of randomness can be controlled by a temperature parameter, which is fine-tuned based on performance on a held-out validation set.
\item \textbf{Diverse Beam Search} \citep{vijayakumar2018diverse}: Diverse Beam Search (DBS) improves conventional beam search by integrating a diversity-augmented objective that explores a wider range of solutions. This method promotes diversity by organizing sequences into multiple groups and applying a diversity penalty during selection, balancing sequence quality and diversity throughout the decoding process. 
\end{enumerate}

\subsubsection{Models}

To demonstrate the effectiveness and adaptability of VPT we applied it to not only our primary LLMC backbone model, CodeT5+, but also to other recent Transformer-based pre-trained code summarization models, as well as a SOTA decoder-only LLMC and a popular online general-purpose LLM.

\begin{enumerate}
\item \textbf{NeuralCodeSum} \citep{NeuralCodeSum}: This enhanced transformer model for code summarization serves as a robust baseline, outperforming various older non-transformer-based models such as CODE-NN \citep{iyer-etal-2016-summarizing}, Tree2Seq \citep{eriguchi-etal-2016-tree}, RL+Hybrid2Seq \citep{rlhybrid}, DeepCom \citep{DeepCom}, API+CODE \citep{transferAPI}, and Dual Mode \citep{DualMode} by a large margin.

\item \textbf{SCRIPT} \citep{SCRIPT}: An advanced transformer-based model that integrates AST information into its inputs, further enhancing its performance on code summarization. 

\item \textbf{PLBART} \citep{ahmad2021PLBART}: A popular BART-based \citep{lewis2019bart} encoder-decoder LLMC that incorporates several pre-training tasks, including both uni-model (code-only) and bi-modal (code+text) tasks prior to fine-tuning on code summarization.

\item \textbf{CodeLlama} \citep{rozière2024codeLLAMA}: CodeLlama is an open-source, decoder-only LLMC that has demonstrated SOTA performance in various code-related tasks, including code generation from natural language \citep{rozière2024codeLLAMA}. We fine-tuned the CodeLlama-7b model using a widely adopted PEFT method, LoRA \citep{hu2022lora}, specifically adapting it for code summarization task. This fine-tuned model serves as a baseline for evaluating how VPT performs relative to other PEFT methods applied to a SOTA decoder-only LLMC.

\item \textbf{GPT-4o} \citep{OpenAI2024GPT4oAPI}: Given the recent popularity of general-purpose LLMs for code-related tasks, we include the popular general-purpose GPT-4o \citep{OpenAI2024GPT4oAPI} as a baseline in our comparisons. Due to limited budget, we compare VPT and GPT-4o on a randomly sampled subset from the test set with a size of 500.

\end{enumerate}

Prior to applying VPT, NeuralCodeSum and SCRIPT model are pre-trained from scratch for our data and CodeT5+ and PLBART are fine-tuned using our data for the code summarization task.

\subsection{Model Settings}
We apply the same parameter settings for the NeuralCodeSum and SCRIPT models, featuring six layers each for the encoder and decoder, and utilizing an eight-head multi-head attention mechanism. The embedding and hidden layers, denoted by \( d_{model} \), have a dimensionality of 512. Both LLMCs, CodeT5+ and PLBART, are equipped with 220 million parameters, including 12 encoder and decoder layers, each with 12 attention heads and a hidden dimension of 768. We use the Adam optimizer \citep{kingma2017adam} for all models with an initial learning rate of 5e-5, no warm-up steps, a batch size of 32, and a dropout rate of 0.2. The models are trained over 200 epochs.

In our experiment, we set the size of the selected subset to \(\#U = \{10, 20\}\). For sampling and VPT, we generate 100 summaries and then apply bi-criteria subset selection to reduce the set to 10 or 20 unique summaries. For beam search, we found that setting the beam size to 100 and applying bi-criteria subset selection leads to a decrease in performance. We therefore opted to generate summaries using beam sizes of 10 or 20 directly. For stochastic and diverse beam search, generating 100 beams simultaneously causes out-of-memory errors. Consequently, for these methods, we directly generate summaries with beam sizes of 10 or 20 and report the results accordingly.

To explore the effectiveness of alternative PEFT methods, we applied one of the most widely adopted PEFT techniques, LoRA Fine-tuning \citep{hu2022lora},  to the CodeLlama model, tailoring it specifically for code summarization tasks. After fine-tuning, we input the source code into the adapted model and sample 100 summaries before applying the bi-criteria subset selection to reduce the set to 10 or 20 unique summaries, ensuring a fair comparison.

For GPT-4o, its closed-source nature prevents the direct integration of VPT into the model. Consequently, we compare its code summarization performance to that of VPT with a CodeT5+ backbone by leveraging the GPT-4o API with sampling. Specifically, following the method proposed by \citeauthor{few-shot-BM25} \citep{few-shot-BM25}, we construct our baseline using BM25 to retrieve the top 10 most similar examples, incorporating each into separate prompts. Similarly, we sample 100 summaries and apply bi-criteria subset selection to generate the desired number of outputs.

The hyperparameters, including the temperature \(T\) for sampling during decoding, the standard deviation scaling factor for sampling latent variables during VPT inference, and the weights for the two criteria in subset selection, are tuned for each dataset based on a held-out validation set.

\subsection{Environment}
Our proposed model is implemented using PyTorch and the Huggingface Transformer Library. Training and inference are performed on a machine equipped with an Intel(R) Core(R) i9-13900KF CPU running at 5.2GHz, 32GB of RAM, and an NVIDIA(R) RTX 4090 GPU.

\begin{table}[ht]
\scriptsize
\centering
\caption{
Performance of VPT compared to other diverse decoding methods: Sampling, Stochastic Beam Search (SBS), Diverse Beam Search (DBS), applied with CodeT5+, evaluated using Oracle BLEU (B), ROUGE-L (R), METEOR (M), BERT-F1 (BT-F1), Precision (BT-PR), and Recall (BT-RE) metrics.
}
\label{tab:diverse_method_compare}

\begin{tabular}{l|l|c|ccccccc}
\hline
\textbf{Lang} & \textbf{Method} & \textbf{\#U} & \textbf{B} & \textbf{R} & \textbf{M} & \textbf{BT-PR} & \textbf{BT-RE} & \textbf{BT-F1} &\textbf{SIDE} \\
\hline

\multirow{10}{*}{Python} 
& Beam           & \multirow{5}{*}{10} & 44.28 & 61.50 & \underline{42.14} & 93.63 & 93.00 & 93.14 & 93.28\\
& Sampling      &                     & \underline{44.31}             & \underline{62.35}             & 41.50             & \underline{93.88}             & \underline{93.19}             & \underline{93.26} & \underline{94.27} \\
& SBS            &                     & 42.30             & 59.76             & 40.50             & 93.25             & 92.76             & 92.80 &  93.33\\
& DBS            &                     & 43.07             & 59.90             & 40.53             & 93.28             & 92.58             & 92.81 & 92.54 \\
& VPT            &                     & \textbf{46.40}    & \textbf{63.83}    & \textbf{43.52}    & \textbf{94.16}    & \textbf{93.38}    & \textbf{93.62} & \textbf{95.21} \\

\cline{2-10}
& Beam           & \multirow{5}{*}{20} & 45.63 & 63.12             & \underline{43.57} & 93.85             & 93.33 & 93.40 & 94.53 \\
& Sampling      &                     & \underline{45.84}          & \underline{64.08} & 42.95         & \underline{94.27} & \underline{93.50}              & \underline{93.58} & \underline{95.16} \\
& SBS            &                     & 43.10             & 60.84             & 41.56             & 93.40             & 93.01             & 93.00 & 94.43\\
& DBS            &                     & 44.49             & 61.81             & 42.76             & 93.64             & 93.14             & 93.17 & 94.56 \\
& VPT            &                     & \textbf{48.62}    & \textbf{66.30}    & \textbf{45.34}    & \textbf{94.75}    & \textbf{93.76}    & \textbf{93.96} & \textbf{96.27} \\
\cline{2-10}
& Beam           & \multirow{3}{*}{100} & 47.13 & 64.70 & 45.37 & 94.01 & 93.71 & 93.59 & 96.10 \\
& Sampling           &   & 48.92 & 67.25 & 45.77 & 94.97 & 93.95 & 94.15 & 95.97 \\

& VPT            &                     & \textbf{51.86}    & \textbf{69.83}    & \textbf{47.75}    & \textbf{95.78}    & \textbf{94.23}    & \textbf{94.60} & \textbf{97.18} \\

\hline
\hline

\multirow{10}{*}{Java} 
& Beam           & \multirow{5}{*}{10} & \underline{47.67} & 62.28 & \underline{45.12} & 93.80 & \underline{93.21} & \underline{93.36} & 91.59\\
& Sampling     &                     & 47.46            & \underline{62.41}             & 44.63             & \underline{93.93}          & 93.12             & 93.34 & \underline{92.42}\\
& SBS            &                     & 46.65             & 61.30             & 44.48             & 93.57             & 93.04             & 93.14 & 91.55 \\
& DBS            &                     & 47.11             & 61.54             & 44.73             & 93.66             & 93.11             & 93.21 & 91.72\\
& VPT            &                     & \textbf{49.22}    & \textbf{64.46}    & \textbf{45.83}    & \textbf{94.52}    & \textbf{93.40}    & \textbf{93.72} & \textbf{93.55} \\
\cline{2-10}
& Beam           & \multirow{5}{*}{20} & \underline{48.11} & 63.07 & \underline{45.81} & 93.90             & \underline{93.42} & \underline{93.49} & 92.72\\
& Sampling      &                     & 48.10             & \underline{63.24}          & 45.22             & \underline{94.10} & 93.28    & \underline{93.49} & \underline{93.09}\\

& SBS            &                     & 46.93             & 61.73             & 46.93             & 93.62             & 93.20             & 93.20 & 92.46 \\
& DBS            &                     & 47.93             & 62.59             & 45.70             & 93.83             & 93.34             & 93.39 & 92.99 \\
& VPT            &                     & \textbf{50.79}    & \textbf{66.12}    & \textbf{47.30}    & \textbf{94.86}    & \textbf{93.78}    & \textbf{94.05} & \textbf{95.02} \\
\cline{2-10}
& Beam           & \multirow{3}{*}{100} & 48.46 & 63.42 & 46.51 & 93.84 & 93.55 & 93.46 & 93.56 \\
& Sampling       &                     & 48.68             & 64.06             & 45.79             & 94.28 & 93.43     & 93.65 & 94.40 \\
& VPT            &                     & \textbf{52.60}    & \textbf{68.01}    & \textbf{48.85}    & \textbf{95.20}    & \textbf{94.11}    & \textbf{94.33} & \textbf{96.47} \\

\hline
\end{tabular}
\end{table}

\section{Results}
This section presents the results of our experiments and provides discussions regarding the RQs.

\subsection{RQ1: Improvement Over Baselines}  
Table~\ref{tab:diverse_method_compare} compares our Variational Prefix-Tuning (VPT) approach with the baseline decoding methods using oracle metrics. VPT consistently outperforms all baselines across both datasets, for both 10 and 20 unique summaries, with more significant improvements observed as \(\#U\) increases. Among the baselines, beam search and sampling yield similar performance, while diverse beam search (DBS) and stochastic beam search (SBS) perform notably worse, suggesting these methods are less suitable for this task. For beam search, sampling, and VPT, we report oracle scores with 100 candidate summaries to provide insight into the upper bound of achievable performance.  VPT demonstrates a substantially higher upper bound on oracle metrics of 100 candidate summaries compared to sampling and beam search. Results for stochastic beam search and diverse beam search are not included due to out-of-memory errors encountered when setting their beam size to 100.

\begin{table}[b]
\centering
\scriptsize
\caption{Diversity of the set of generated summaries using different diverse decoding methods compared to VPT, applied with CodeT5+, measured by Distinct Unigram (D-1) and Bigram (D-2) in percentage and Self-BLEU(S-B) scores.}
\label{tab:diversity performance}
\begin{tabular}{l|c|ccc|ccc}
\hline
\multirow{2}{*}{Method} & \multirow{2}{*}{\#U} & \multicolumn{3}{c|}{Python} & \multicolumn{3}{c}{Java} \\ \cline{3-8}
                &      & D-1↑ & D-2↑ & S-B↓ & D-1↑ & D-2↑ & S-B↓\\
\hline
Beam           &  & 16.79	& 33.02	& 79.63	& 15.15	& 27.40	& 82.04	 \\
Sampling       &    & \textbf{38.80}	& \textbf{66.49}	& \textbf{48.52}	& \textbf{56.08}	& \textbf{72.52}	& \underline{72.59}	  \\
SBS            & 10   & 21.33	& 36.45	& 77.41	& 19.46	& 31.33 & 80.43	  \\
DBS            &    & 26.84	& 43.10	& 76.91	& 10.95	& 18.88	& 82.80	 \\
VPT            &  & \underline{28.36}	& \underline{51.97}	& \underline{59.05} & \underline{30.92} & \underline{52.81} & \textbf{56.83}  \\
\hline\hline
Beam           &  & 10.64	& 23.45	& 84.18	& 9.29	& 18.60	& 85.87	  \\
Sampling       &    & \textbf{31.90}	& \textbf{58.69}	& \textbf{54.98}	& \textbf{52.28}& \textbf{67.91}	& \underline{76.46}	 \\
SBS            & 20   & 13.61	& 26.25	& 83.15 & 12.62	& 22.72	& 85.03\\
DBS            &    & 11.95	& 21.54	& 86.20	& 21.05	& 31.85	& 87.70	 \\

VPT            &  & \underline{20.82}	& \underline{42.91}	& \underline{64.48} & \underline{23.40}	& \underline{44.38}	& \textbf{62.13}	  \\
\hline

\end{tabular}

\end{table}

\paragraph{RQ1 Conclusion}  
The empirical results indicate that the set of summaries generated by VPT is more likely to include an accurate summary compared to the baselines, suggesting that our approach is more apt to provide an adequate summary in practical use. Moreover, as the number of unique summaries ($\#U$) increases, the performance gains of VPT become even more pronounced, demonstrating that increased diversity enhances the likelihood of generating a more accurate summary. Additionally, the higher upper bound of VPT’s 100-sample oracle scores points to further potential improvements if more advanced subset selection algorithms are implemented.

\begin{table}[H]
\centering
\scriptsize
\caption{Qualitative example for demonstrating diversity of different methods}
\label{tab:qualitative_example_python}
\begin{tabular}{@{}p{\textwidth}@{}} 
\toprule

\begin{minipage}{\textwidth}
\textbf{Source code:} 
\begin{lstlisting}
def default_selem(func):
  @functools.wraps(func)
  def func_out(image, selem=None, *args, **kwargs):
    if selem is None:
      selem = _default_selem(image.ndim)
    return func(image, selem=selem, *args, 
                **kwargs)
  return func_out
\end{lstlisting} 
\end{minipage}\\
\midrule
\textbf{Variational Prefix Tuning:} \\
1. decorator to add a default structuring element to morphological operations.\\
2. generates a binary structure for the given selem.\\
3. decorator to generate a binary structure of the image if no selem is specified.\\
4. return a decorator that generates a binary structure for the given selem.\\
5. return a binary structure for the given selem.\\
6. decorator to generate a default structuring element for morphological operations.\\
\midrule
\textbf{Beam Search Generation:} \\
1. decorator to add a default structuring element to morphological operations.\\
2. decorator to add a default structuring element to morphology functions.\\
3. decorator to generate a binary structure for binary morphological operations.\\
4. decorator to generate a binary structure for an image if no selem is specified.\\
5. decorator to generate a binary structuring element if no selem is specified.\\
6. decorator to generate a binary structure for morphological operations.\\
\midrule
\textbf{Sampling Generation:} \\
1. decorator to give a non-standard "selem" field a binary structure of the outer dimensions of the image.\\
2. decorator to create a classical binary structure for image parameters func : function the function to create.\\
3. decorator to provide a default structure for grayscale functions.\\
4. decorator to generate a binary structure for features without an selem.\\
5. decorator to add a default structuring element to binary structuring elements of a binary image .\\
6. decorator to generate a binary structure for grayscale functions.\\
\midrule
\textbf{Diverse Beam Search Generation:} \\
1. decorator to add a default structuring element to morphological operations.\\
2. decorator to add a default structuring element to morphology functions.\\
3. decorator to generate a binary structure for binary morphological operations.\\
4. decorator to add a default structuring element to binary image.\\
5. decorator to add a default structuring element to binary image structure if no selem is specified.\\
6. decorator to add a default structuring element to morphological operations.\\
\midrule
\textbf{Stochastic Beam Search Generation:} \\
1. decorator to generate a binary structure for binary morphological operations.\\
2. decorator to generate a binary structure for an image if no selem is specified.\\
3. decorator to generate a binary structure for sub-images of the same shape as the input image.\\
4. decorator to generate a binary structure for morphological operations.\\
5. decorator to generate a binary structure for inputs with the same shape as the given selem.\\
6. decorator to generate a binary structure for sub-image functions.\\
\bottomrule
\end{tabular}
\end{table}

\subsection{RQ2: Diversity of Generated Summaries}
In this section, we analyze the diversity of generated summaries from VPT compared to other generation methods, examining both quantitative and qualitative aspects to gain a comprehensive understanding of the results.

\paragraph{Diversity Analysis}
Table~\ref{tab:diversity performance} presents Distinct N-grams and Self-BLEU scores to evaluate the diversity of the generated outputs from baseline generation methods and VPT. We observe that, across all diversity metrics, VPT ranks second, just below the sampling method, which is the most prevalent method used for generating diverse results in LLMs. Beam search, in contrast, consistently results in the lowest diversity scores across most settings. While its diversity-enhanced variants, stochastic and diverse beam search, show some improvement, they still fall short of the diversity achieved by VPT.

\paragraph{Qualitative Examples}
Table~\ref{tab:qualitative_example_python} presents a comparative analysis of results produced by the baseline methods and VPT applied to CodeT5+ for a single source code from the Python dataset. These qualitative examples provide insight into the diversity of each method's generated outputs.

\paragraph{RQ2 Conclusion}
The comparative results indicate that our method generates a more diverse set of solutions compared to beam search and its two diversity-enhancing variants, thereby increasing the likelihood that the generated set includes an accurate summary. Although sampling generally achieves greater diversity than VPT—as demonstrated in RQ1—it consistently exhibits a lower probability of containing an accurate summary in the generated outputs. This suggests that while diversity is beneficial, excessive diversity may compromise the overall quality of the summaries.

\subsection{RQ3: Performance of VPT When Applied to Other Pre-trained Models and Compared to Other Models}

\paragraph{Effectiveness on Other Models} Table~\ref{tab:model-performance} presents the performance of the top three decoding methods based on our experiments with CodeT5+. The results demonstrate that VPT significantly enhances performance across all pre-trained models and both \(\#U\) values for every evaluation metric and dataset. The degree of improvement varies by dataset, metric, and \(\#U\), with enhancements at \(\#U=20\) being more pronounced than at \(\#U=10\).
\begin{table}[H]
\centering
\setlength{\tabcolsep}{4pt}
\scriptsize
\caption{Performance of VPT applied with other pre-trained models.}
\label{tab:model-performance}
\begin{tabular}{l|l|c|ccccccc}
\hline
\textbf{Lang} & \textbf{Model} & \textbf{\#U} & \textbf{B} & \textbf{R} & \textbf{M} & \textbf{BT-PR} & \textbf{BT-RE} & \textbf{BT-F1}  & \textbf{SIDE}\\
\hline

\multirow{24}{*}{Python} 
& NeuralCodeSum w/ Beam    & \multirow{13}{*}{10} & 38.77 & 55.14 & 33.96 & 92.70 & 91.27 & 91.75 & 90.87\\
& NeuralCodeSum w/ Sample  &  & 39.13 & 55.71 & 34.81 & 92.52 & 91.61 & 91.81 &  91.19\\
& NeuralCodeSum w/ VPT     &  & \textbf{41.44} & \textbf{57.83} & \textbf{36.51} & \textbf{93.20} & \textbf{91.98} & \textbf{92.32} & \textbf{92.19}\\
\cline{2-2}\cline{4-10}
& SCRIPT w/ Beam           &  & 40.15 & 56.46 & 34.87 & 93.05 & 91.44 & 92.01 & 83.18 \\
& SCRIPT w/ Sample         &  &  40.59 & 57.11 & 35.88 & 92.90 & 91.84 & 92.11 & 83.74 \\
& SCRIPT w/ VPT            &  & \textbf{42.54} & \textbf{58.96} & \textbf{37.15} & \textbf{93.40} & \textbf{92.08} & \textbf{92.49} & \textbf{85.09}\\
\cline{2-2}\cline{4-10}
& PLBART w/ Beam           &  & 42.46 & 58.92 & 39.76 & 93.12 & 92.54 & 92.63 & 92.82 \\
& PLBART w/ Sample         &  & 42.95 &  59.74 & 39.55 & 93.46 & 92.59 & 92.77 & 93.25 \\
& PLBART w/ VPT            &  & \textbf{44.20} & \textbf{61.03} & \textbf{40.44 }& \textbf{93.92} & \textbf{92.84} & \textbf{93.09} & \textbf{94.14}\\
\cline{2-2}\cline{4-10}
& CodeT5+ w/ Beam          &  & 44.28 & 61.50 & 42.14 & 93.63 & 93.00 & 93.14 & 93.28\\
& CodeT5+ w/ Sample        &  & 44.31 & 62.35 & 41.50 & 93.88 & 93.19 & 93.26 & 94.27\\
& CodeT5+ w/ VPT           &  & \textbf{46.40} & \textbf{63.83} & \textbf{43.52} & \textbf{94.16} & \textbf{93.38} & \textbf{93.62} & \textbf{95.21}\\
\cline{2-10}
& NeuralCodeSum w/ Beam    & \multirow{13}{*}{20} & 40.08 & 57.05 & 35.25 & 93.19 & 91.60 & 92.11 & 92.93\\
& NeuralCodeSum w/ Sample  &  & 40.05 & 57.10 & 35.74 & 92.88 & 91.85 & 92.09 & 92.51\\
& NeuralCodeSum w/ VPT     &  & \textbf{42.89} & \textbf{59.77} & \textbf{37.68} & \textbf{93.69} & \textbf{92.27} & \textbf{92.65} & \textbf{93.91} \\
\cline{2-2}\cline{4-10}
& SCRIPT w/ Beam           &  & 41.47 & 58.04 & 36.17 & 93.55 & 91.77 & 92.37 & 86.26 \\
& SCRIPT w/ Sample         &  & 41.55 & 58.51 & 36.78 & 93.24 & 92.08 & 92.37 &  86.59 \\
& SCRIPT w/ VPT            &  & \textbf{43.88} & \textbf{60.80} & \textbf{38.39} & \textbf{93.91} & \textbf{92.36} & \textbf{92.86} & \textbf{88.29}\\
\cline{2-2}\cline{4-10}
& PLBART w/ Beam           &  & 43.56 & 60.34 & 41.06 & 93.35 & 92.85 & 92.87 & 94.16\\
& PLBART w/ Sample         &  & 43.92 & 61.07 & 40.50 & 93.75 & 92.82 & 93.02 & 94.39\\
& PLBART w/ VPT            &  & \textbf{46.17} & \textbf{63.37} & \textbf{42.33} & \textbf{94.40} & \textbf{93.24} & \textbf{93.49} & \textbf{95.41}\\
\cline{2-2}\cline{4-10}
& CodeT5+ w/ Beam          &  & 45.63 & 63.12 & 43.57 & 93.85 & 93.33 & 93.40 & 94.53\\
& CodeT5+ w/ Sample        &  & 45.84 & 64.08 & 42.95 & 94.27 & 93.50 & 93.58 & 95.16\\ 
& CodeT5+ w/ VPT           &  & \textbf{48.62} & \textbf{66.30} & \textbf{45.34} & \textbf{94.75} & \textbf{93.76} & \textbf{93.96} &\textbf{96.27} \\
\hline
\hline

\multirow{24}{*}{Java} 
& NeuralCodeSum w/ Beam    & \multirow{13}{*}{10} & 47.40 & 59.25 & 45.13 & 93.26 & 91.99 & 92.44 & 92.18\\
& NeuralCodeSum w/ Sample  &  & 49.04 & 60.90 & 47.11 & 93.41 & 92.60 & 92.76 & 91.47 \\
& NeuralCodeSum w/ VPT     &  & \textbf{51.02} & \textbf{62.42} & \textbf{48.67}  & \textbf{93.93} & \textbf{92.87} & \textbf{93.17} & \textbf{93.41} \\
\cline{2-2}\cline{4-10}
& SCRIPT w/ Beam           &  & 50.70 & 62.48 & 48.39 & 94.10 & 92.62 & 93.17 &  91.25\\
& SCRIPT w/ Sample         &  & 51.62 & 63.59 & 49.70 & 94.15 & 93.09 & 93.37 & 92.17\\
& SCRIPT w/ VPT            &  & \textbf{53.01} & \textbf{64.73} & \textbf{50.30} & \textbf{94.64} & \textbf{93.17} & \textbf{93.71} & \textbf{92.55}\\
\cline{2-2}\cline{4-10}
& PLBART w/ Beam           &  & 46.13 & 60.45 & 43.67 & 93.32 & 92.91 & 92.95 & 91.08\\
& PLBART w/ Sample         &  & 46.30 & 61.12 & 43.40 & 93.67 & 92.83 & 93.06 & 91.67 \\
& PLBART w/ VPT            &  & \textbf{47.21} & \textbf{62.30} & \textbf{44.38} & \textbf{93.94} & \textbf{93.10} & \textbf{93.29} & \textbf{92.92}\\
\cline
{2-2}\cline{4-10}
& CodeT5+ w/ Beam          &  & 47.67 & 62.28 & 45.12 & 93.80 & 93.21 & 93.36 & 91.59 \\
& CodeT5+ w/ Sample        &  & 47.46 & 62.41 & 44.63 & 93.93 & 93.12 & 93.34 & 92.42 \\
& CodeT5+ w/ VPT           &  & \textbf{49.22} & \textbf{64.46} & \textbf{45.83} & \textbf{94.52} & \textbf{93.40} & \textbf{93.72} & \textbf{93.55}\\

\cline{2-10}
& NeuralCodeSum w/ Beam    & \multirow{13}{*}{20} & 48.21 & 60.56 & 45.89 & 93.61 & 92.19 & 92.68 & 93.78\\
& NeuralCodeSum w/ Sample  &  & 49.84 & 61.97 & 48.06 & 93.68 & 92.80 & 92.97 & 93.16\\
& NeuralCodeSum w/ VPT     &  & \textbf{51.85} & \textbf{63.68} & \textbf{49.49} & \textbf{94.29} & \textbf{93.01} & \textbf{93.44} & \textbf{94.66}\\
\cline{2-2}\cline{4-10}
& SCRIPT w/ Beam           &  & 51.62 & 63.81 & 49.25 & 94.43 & 92.82 & 93.41 & 93.12\\
& SCRIPT w/ Sample         &  & 52.41 & 64.69 & 50.47 & 94.41 & 93.28 & 93.57 & 93.32\\
& SCRIPT w/ VPT            &  & \textbf{53.90} & \textbf{66.10} & \textbf{51.12} & \textbf{94.93} & \textbf{93.36} & \textbf{93.94} & \textbf{94.00}\\
\cline{2-2}\cline{4-10}
& PLBART w/ Beam           &  & 46.64 & 61.15 & 44.38 & 93.38 & 93.09 & 93.04 & 92.02\\
& PLBART w/ Sample         &  & 47.12 & 62.17 & 44.19 & 93.88 & 93.05 & 93.24 & 92.60 \\
& PLBART w/ VPT            &  & \textbf{48.81} & \textbf{64.08} & \textbf{45.77} & \textbf{94.33} & \textbf{93.42} & \textbf{93.60} & \textbf{94.25} \\
\cline{2-2}\cline{4-10}
& CodeT5+ w/ Beam          &  & 48.11 & 63.07 & 45.81 & 93.90 & 93.42 & 93.49 & 92.72 \\
& CodeT5+ w/ Sample        &  & 48.10 & 63.24 & 45.22 & 94.10 & 93.28 & 93.49 & 93.09 \\
& CodeT5+ w/ VPT           &  & \textbf{50.79} & \textbf{66.12} & \textbf{47.30} & \textbf{94.86} & \textbf{93.78} & \textbf{94.05} & 
\textbf{95.02}\\

\hline
\end{tabular}
\end{table}

\begin{table}[htbp!]
\centering
    \scriptsize
    \setlength{\tabcolsep}{4pt}
    \caption{Performance of Lora-finetuned CodeLlama model compared to VPT on full test set.}
    \label{tab:CodeLlama compare}
    \begin{tabular}{l|l|c|ccccccc}
      \hline
      \textbf{Lang} & \textbf{Model} & \textbf{\#U} & \textbf{B} & \textbf{R} & \textbf{M} & \textbf{BT-PR} & \textbf{BT-RE} & \textbf{BT-F1} & \textbf{SIDE} \\
      \hline
      \multirow{4}{*}{Python} 
      & CodeLlama w/ Lora          & \multirow{2}{*}{10}  & 32.86 & 54.99  & 29.11  & 93.27 & 91.46 & 92.02 & 94.95 \\
      & VPT                        &                      & \textbf{46.40} & \textbf{63.83} & \textbf{43.52} & \textbf{94.16} & \textbf{93.38} & \textbf{93.62} & \textbf{95.21} \\
      \cline{2-10}
      & CodeLlama w/ Lora          & \multirow{2}{*}{20}  & 35.41 & 57.92 & 31.41 & 93.86 & 92.05 & 92.53 & 95.61 \\
      & VPT                        &                      & \textbf{48.62} & \textbf{66.30} & \textbf{45.34} & \textbf{94.75} & \textbf{93.76} & \textbf{93.96} & \textbf{96.27} \\
      \hline\hline
      \multirow{4}{*}{Java} 
      & CodeLlama w/ Lora          & \multirow{2}{*}{10}  & 27.06 & 47.79 & 25.89 & 92.47 & 90.09 & 90.85 & \textbf{94.43} \\
      & VPT                        &                      & \textbf{49.22} & \textbf{64.46} & \textbf{45.83} & \textbf{94.52} & \textbf{93.40} & \textbf{93.72} & 93.55 \\
      \cline{2-10}
      & CodeLlama w/ Lora          & \multirow{2}{*}{20}  & 29.45 & 50.52 & 28.09 & 92.98 & 90.68 & 91.29 & \textbf{95.63} \\
      & VPT                        &                      & \textbf{50.79} & \textbf{66.12} & \textbf{47.30} & \textbf{94.86} & \textbf{93.78} & \textbf{94.05} & 95.02 \\
      \cline{2-9}
      \hline
    \end{tabular}

\end{table}

\paragraph{Comparison to LoRA-finetuned CodeLlama} To provide a comparison with other PEFT methods, we compare our approach to a  SOTA decoder-only model, CodeLlama, specialized for code summarization using the widely adopted PEFT method LoRA. Our results in Table~\ref{tab:CodeLlama compare} indicate that our approach significantly outperforms the LoRA-finetuned CodeLlama across most metrics, with the only exception being the SIDE score on the Java dataset. These findings suggest that our method is comparable to, or even outperforms, other PEFT techniques.

\paragraph{Comparison to GPT-4o} For the general-purpose LLM GPT-4o, experiments were conducted on a subset of 500 examples due to budget constraints, with results presented in Table~\ref{tab:GPT4o_on_subset_compare}. Across all metrics, our approach consistently outperforms GPT-4o with few-shot learning using BM25 example retrieval as proposed by \citeauthor{few-shot-BM25} \citep{few-shot-BM25} by a significant margin. These results confirm that our method can match or exceed the performance of a general-purpose model enhanced with state-of-the-art few-shot learning techniques.

\begin{table}[htbp!]
\scriptsize
\setlength{\tabcolsep}{4pt}
\centering
\caption{
Performance of GPT4o compared to VPT on a 500 samples from the test set.
}

\label{tab:GPT4o_on_subset_compare}

\begin{tabular}{l|l|c|ccccccc}
\hline
\textbf{Lang} & \textbf{Model} & \textbf{\#U} & \textbf{B} & \textbf{R} & \textbf{M} & \textbf{BT-PR} & \textbf{BT-RE} & \textbf{BT-F1} & \textbf{SIDE} \\
\hline

\multirow{4}{*}{Python} 
& GPT-4o w/ BM-25           & \multirow{2}{*}{10} & 36.51 & 50.54 &  41.62 &  92.22 &  91.51 &  91.70 & 93.99   \\
& VPT            &          & \textbf{45.45} & \textbf{62.30} & \textbf{42.00} & \textbf{93.88} & \textbf{93.19} & \textbf{93.26} & \textbf{95.09}\\

\cline{2-10}
& GPT-4o w/ BM-25           &  \multirow{2}{*}{20}  & 37.55 & 52.02 & 42.57 & 92.45 &  91.70 &  91.89 & 94.67\\
& VPT            &          & \textbf{47.53} & \textbf{64.59} & \textbf{43.70} & \textbf{94.43} & \textbf{93.51} & \textbf{93.68} & \textbf{95.70}\\

\hline
\hline

\multirow{4}{*}{Java} 
& GPT-4o w/ BM-25          & \multirow{2}{*}{10}  & 32.83 & 46.21 & 36.41 & 91.10 & 91.41 & 91.02 & 92.95 \\
& VPT            &         & \textbf{49.34} & \textbf{64.46} & \textbf{45.76} & \textbf{94.44} & \textbf{93.30} & \textbf{93.64} & \textbf{93.52}\\
\cline{2-10}
& GPT-4o w/ BM-25          & \multirow{2}{*}{20}   & 34.51 & 49.17 & 37.70 & 91.54 & 91.87 & 91.43 & 93.54 \\
& VPT            &         & \textbf{50.74} & \textbf{65.74} & \textbf{46.98} & \textbf{94.79} & \textbf{93.56} & \textbf{93.93} & \textbf{94.80}\\
\cline{2-9}

\hline
\end{tabular}
\end{table}

\paragraph{RQ3 Conclusion} Overall, our findings highlight the effectiveness and adaptability of VPT when applied to various models. Specifically, the application of VPT to CodeT5+ outperforms both the LoRA-finetuned, decoder-only CodeLlama and the general-purpose GPT-4o enhanced with few-shot learning \citep{few-shot-BM25}.

\subsection{RQ4: Ablation Study}

In this section, we evaluate the contributions of three additional components that enhance the VPT method, as detailed in Table~\ref{tab:Ablation study}. To evaluate the statistical significance of the observed contributions, we performed a one-sided Wilcoxon signed-rank test, with the alternative hypothesis asserting that the median difference between paired observations is greater than zero. We compared the metric values before and after applying each component and report the corresponding p-values in Table~\ref{tab:wilcoxon}.

\paragraph{Beam Search} We compare beam search decoding (with a beam size of 4) against greedy decoding for each sampled latent variable's decoding process. For the Python dataset, beam search improves BLEU and METEOR scores by 1.3 and 1.1 points, respectively, and ROUGE scores by 1.1 and 0.9 points for \(\#U = 10\) and \(\#U = 20\), respectively. Additionally, semantic similarity-based metrics such as BERT scores and SIDE score improve by approximately 0.15 points. For the Java dataset, we observe the BLEU, ROUGE, and METEOR scores increase by 0.5–0.6 points for $\#U = 10$ and by 0.3–0.4 points for $\#U = 20$, and an approximate 0.15-point enhancement observed for all semantic similarity-based metrics (BERT and SIDE score).

\paragraph{Prior Net} As discussed in Section~\ref{subsec:Prior}, we parameterize the prior distribution using the source code contextual embedding generated by the CodeT5+ encoder, rather than relying on the standard Gaussian prior \(\mathcal{N}(0,1)\) commonly employed in previous VAE/CVAE approaches \citep{wang2017diverseandaccurate, CreativityDiverseQuestion, CVAE}. Sampling from the prior distribution generated by our prior network results in consistent—albeit modest—improvements across all metrics, \(\#U\) values, and datasets, yielding approximately a 0.3-point increase in BLEU, ROUGE-L, and METEOR scores, along with gains of 0.1–0.15 points in BERT and SIDE scores.

\paragraph{Bi-criteria Subset Selection} Incorporating bi-criteria subset selection, rather than simply taking the first \(\#U\) unique summaries generated, leads to substantial performance improvements. For the Python dataset, this strategy results in increases of 1.3 and 1.0 points in BLEU and METEOR scores for \(\#U = 10\) and \(\#U = 20\), respectively, while ROUGE scores improve by 0.8 and 0.7 points. Additionally, BERT-recall and F1 scores show enhancements of 0.2 and 0.1 points. For the Java dataset, BLEU scores improve by approximately 0.9 points for both $\#U = 10$ and $\#U = 20$, with ROUGE scores increasing by 0.7 and 0.9 points and METEOR scores by 0.5 points. Semantic similarity metrics, including BERT and SIDE scores, exhibit improvements of roughly 0.1–0.2 points across all settings.
\begin{table}[htbp!]
\centering
\scriptsize
\caption{Ablation study on contribution to the overall performance of Beam Search(BS), Prior Net (PN), and Bi-criteria Subset Selection (BC) on Python and Java datasets.}
\label{tab:Ablation study}

\begin{tabular}{l|ccc|c|ccccccc}
\hline
Lang &BS & PN & BC  &   \#U    & B     & R     & M     & BT-PR & BT-RE & BT-F1 & SIDE \\ 
\hline
\multirow{8}{*}{python} & \multirow{2}{*}{$\times$} & \multirow{2}{*}{$\times$} & \multirow{2}{*}{$\times$}        
        & 10  & 43.45 & 61.51 & 40.72 & 93.78 & 92.96 & 93.13 & 94.40\\
& &   &   & 20  & 46.28 & 64.51 & 43.07 & 94.44 & 93.43 & 93.65 & 95.60\\

\cline{2-12}

& \multirow{2}{*}{\checkmark} & \multirow{2}{*}{$\times$} & \multirow{2}{*}{$\times$}
        & 10  & 44.88 & 62.68 & 42.06 & 93.98 & 93.12 & 93.32 & 94.54\\
& &   &   & 20  & 47.43 & 65.49 & 44.04 & 94.65 & 93.58 & 93.81 & 95.74\\
\cline{2-12}

& \multirow{2}{*}{\checkmark} & \multirow{2}{*}{\checkmark} & \multirow{2}{*}{$\times$} 
        & 10  & 45.13 & 63.00 & 42.26 & 94.10 & 93.16 & 93.40 & 94.62\\
& &   &   & 20  & 47.71 & 65.75 & 44.32 & 94.75 & 93.67 & 93.89 & 95.89 \\
\cline{2-12}

& \multirow{2}{*}{\checkmark} & \multirow{2}{*}{\checkmark} & \multirow{2}{*}{\checkmark} 
        & 10  & 46.40 & 63.83 & 43.52 & 94.16 & 93.38 & 93.62 & 95.11\\
& &   &   & 20  & 48.62 & 66.30 & 45.34 & 94.75 & 93.76 & 93.96 & 96.21\\

\hline
\hline
\multirow{8}{*}{java} & \multirow{2}{*}{$\times$} & \multirow{2}{*}{$\times$} & \multirow{2}{*}{$\times$}        
        & 10  & 47.13 & 62.37 & 44.31 & 93.95 & 93.14 & 93.30 & 92.72 \\
& &   &   & 20  & 49.24 & 64.50 &46.21 & 94.42 & 93.51 & 93.70 & 94.36 \\

\cline{2-12}
& \multirow{2}{*}{\checkmark} & \multirow{2}{*}{$\times$} & \multirow{2}{*}{$\times$}
        & 10  & 47.72 & 62.86 & 44.79 & 94.05 &  93.27 &  93.43 & 92.82\\
& &   &   & 20  & 49.59 &  64.90 &  46.49 &  94.47 &  93.60 &  93.78 &  94.43\\
\cline{2-12}
& \multirow{2}{*}{\checkmark} & \multirow{2}{*}{\checkmark} & \multirow{2}{*}{$\times$} 
        & 10  & 48.24 & 63.77 & 45.33 & 94.17 & 93.33 & 93.56 & 93.06 \\
& &   &   & 20  & 49.93 &  65.22 & 46.80 & 94.55 & 93.64 & 93.84 & 94.58 \\
\cline{2-12}

& \multirow{2}{*}{\checkmark} & \multirow{2}{*}{\checkmark} & \multirow{2}{*}{\checkmark} 
        & 10  & 49.22 &  64.46 &  45.83 &  94.52 &  93.40 &  93.72 & 93.55 \\
& &   &   & 20  & 50.79 & 66.12 & 47.30 & 94.86 & 93.78 & 94.05 & 95.02 \\ 

\hline

\end{tabular}
\end{table}
\paragraph{Wilcoxon Test} We set the significance level $\alpha$ for the Wilcoxon signed-rank test at 0.05. Our results presented in Table~\ref{tab:wilcoxon} indicate that, for nearly all metrics, the improvements from adding each component are statistically significant. The few exceptions are as follows: for the Python dataset, the improvements in the SIDE and METEOR metrics when incorporating the Prior Net component for generating 10 summaries, as well as the improvement in BERT-precision when applying Bi-criteria re-ranking for generating 20 summaries, did not reach significance. For the Java dataset, the exceptions include the SIDE improvement when adding Beam Search for generating 10 summaries and the BERT-precision improvement when integrating the Prior Net. Overall, these findings confirm that most of the enhancements observed for each individual component are statistically significant.
\begin{table}[htbp!]
\centering

    \setlength{\tabcolsep}{5.5pt}
    \tiny
    \caption{Wilcoxon signed-rank test p-value results for the ablation study. Comparisons are made between the model's performance with and without the additional components.}
    \label{tab:wilcoxon}
    \begin{tabular}{l|ccc|c|ccccccc}
    \hline
    Lang & BS & PN & BC  & \#U    & B     & R     & M     & BT-PR & BT-RE & BT-F1 & SIDE \\ 
    \hline
    \multirow{6}{*}{python} 
& \multirow{2}{*}{\checkmark} & \multirow{2}{*}{$\times$} & \multirow{2}{*}{$\times$}
        & 10  & 7.79e-34 & 5.64e-40 & 3.83e-34 & 4.98e-34 & 1.81e-21 & 5.40e-36 & 3.46e-06\\
& &   &   & 20  & 8.31e-29 & 4.90e-33 & 2.58e-21 & 5.68e-22 & 3.72e-11 & 1.76e-26 & 1.76e-26\\
\cline{2-12}

& \multirow{2}{*}{\checkmark} & \multirow{2}{*}{\checkmark} & \multirow{2}{*}{$\times$} 
        & 10  & 1.41e-04 & 3.72e-05 & 0.17 & 4.51e-04 & 4.4e-02 & 1.5e-04 & 0.11\\
& &   &   & 20  & 2.86e-14 & 1.30e-10 & 1.07e-11 & 8.56e-12 & 5.09e-13 & 1.21e-09 & 4.11e-02 \\
\cline{2-12}

& \multirow{2}{*}{\checkmark} & \multirow{2}{*}{\checkmark} & \multirow{2}{*}{\checkmark} 
        & 10  & 1.09e-44 & 1.11e-30 & 1.19e-80 & 1.59e-05 & 6.41e-62 & 2.09e-19 & 9.38e-73\\
& &   &   & 20  & 2.07e-12 & 3.03e-10 & 2.53e-37 & 0.90 & 1.46e-44 & 6.84e-09 & 4.06e-49\\ 
\hline
\hline
\multirow{6}{*}{java} 
& \multirow{2}{*}{\checkmark} & \multirow{2}{*}{$\times$} & \multirow{2}{*}{$\times$}
        & 10  & 8.37e-06 & 3.75e-07 & 6.98e-08 & 6.00e-15 &  3.01e-09 &  1.11e-14 & 0.475\\
& &   &   & 20  & 5.62e-07 &  1.94e-09 &  1.21e-05 &  2.15e-17 &  6.90e-3 &  2.39e-09 &  1.19e-3\\
\cline{2-12}
& \multirow{2}{*}{\checkmark} & \multirow{2}{*}{\checkmark} & \multirow{2}{*}{$\times$} 
        & 10  & 2.81e-09 & 4.06e-10 & 5.06e-11 & 8.10e-07 & 1.93e-07 &  5.92e-14 & 9.99e-08 \\
& &   &   & 20  & 2.00e-03 &  7.6e-03 & 1.46e-06 & 0.32 & 3.37e-08 & 9.53e-08 & 5.71e-10 \\
\cline{2-12}

& \multirow{2}{*}{\checkmark} & \multirow{2}{*}{\checkmark} & \multirow{2}{*}{\checkmark} 
        & 10  & 2.26e-26 &  2.14e-44 &  7.71e-06 &  2.48e-95 &  1.16e-15 & 7.92e-60 & 9.97e-17 \\
& &   &   & 20  & 4.59e-25 & 2.88e-40 & 3.22e-09 & 1.09e-95 & 5.99e-18 & 4.94e-64 & 1.60e-16 \\ 
    \hline
    \end{tabular}

\end{table}
\paragraph{RQ4 Conclusion} The ablation study demonstrates the effectiveness of the three components used to enhance the VPT method. The integration of Beam Search, Prior Net, and Bi-criteria Subset Selection yields considerable improvements in performance for both the Python and Java datasets. Furthermore, the statistical significance of these improvements, as confirmed by the Wilcoxon signed-rank test, validates the contributions of each component.

\section{Threats to Validity}

\paragraph{Dataset Limitations}
The datasets utilized in our study present specific challenges that could impact the validity of our findings. These include the presence of poorly written code summaries and the limitation to specific programming languages. The lack of dataset diversity, particularly in terms of programming languages, raises questions about whether the observed effectiveness of our approach extends to other languages. Although we included two widely used datasets for code summarization, encompassing two different programming languages to better illustrate our model's capabilities, expanding the dataset to additional languages for a more comprehensive evaluation is a direction for future work.

\paragraph{Limitations of Evaluation Metrics}
The validity of our model's evaluation might be influenced by our reliance on lexical metrics such as BLEU, ROUGE-L, and METEOR, which may not entirely capture the subtleties of generated code summaries. While the BERTScore is designed to better capture semantic similarity, it's possible that this metric, along with the metrics used for measuring diversity, may not fully reflect the actual accuracy and diversity of the summaries. Therefore, there's a need for future research to explore more robust metrics that can accurately gauge both the semantic quality and the diversity of the summaries. Incorporating human judgment into the evaluation process, despite potential subjective biases, could also provide invaluable insights into the practical utility and readability of the summaries.

\paragraph{Reliance on Human Judgment}
While we have demonstrated that our method can increase the likelihood of producing a more accurate summary within a set of diverse outputs, the ground truth summaries we use as references are collected from real developer summaries. These human-written summaries can be subject to individual biases and can vary widely due to different interpretations and preferences among developers. Consequently, the effectiveness of our model may differ in real-world scenarios. For our current work, we adopted Oracle metrics to provide an upper bound on performance, reflecting an ideal selection process. In future work, we plan to include human studies to provide more comprehensive evaluations of overall performance.

\section{Related Work}

\subsection{Code Summarization}

Early studies in code summarization approached the task as an information retrieval (IR) problem, concentrating on extracting keywords from source code to form natural language descriptions \citep{IR-based-CS1, IR-based-CS2, IR-based-CS3}. However, these methods often yielded summaries with limited readability, failing to capture the essential semantic relationships between source code and natural language \citep{CodeSumSurvey}.

The advent of deep learning and the availability of large-scale data brought a shift towards seq2seq models, particularly with the use of RNNs. Pioneering works utilized LSTM with attention mechanisms to create source code summaries \citep{iyer-etal-2016-summarizing}. Despite their initial success, RNNs struggle with long-range dependencies in source code due to their sequential nature. The Transformer model \citep{transformer} addressed these limitations with a self-attention mechanism, leading to significant improvements in various code-related generation tasks, including code summarization \citep{NeuralCodeSum}.

The integration of structural information, such as Abstract Syntax Trees (ASTs), into the summarization process represents another significant advancement. Approaches like training LSTM models on AST sequences flattened using Structure-Based Traversal (SBT) \citep{DeepCom} and techniques that combine text and structural representations \citep{LeClair_etal_TextSBTCombined} have enhanced the learning of code semantics. Moreover, incorporating structural information into Transformer-based models \citep{SCRIPT, SG-transformer, wu-etal-sit-transformer} has further improved the quality of generated summaries. Additionally, models such as Graph Neural Networks (GNNs) \citep{LeClair_GNN}, Convolutional Neural Networks (CNNs) \citep{Yu_etal_cnn}, and Graph Attention Networks (GATs) \citep{GATCodeSum}, which encode structural information within code, have also demonstrated potential in code summarization.

\subsection{Diversity in Generation}
Advancements in natural language processing models have greatly improved the accuracy of generated code summaries. Concurrently, there has been a growing focus on diverse generation within natural language processing, a key aspect in creating contextually varied text, as explored in recent studies \citep{tevet-berant-2021-evaluating, holtzman2020curious}.

One major approach to improving diversity involves incorporating generative models. For instance, applying adversarial training to seq2seq models, as proposed by \citep{li-etal-2018-generating, xu-etal-2018-diversity}, leverages Generative Adversarial Networks (GANs) \citep{goodfellow2014generative} for diverse text or story generation. VAE and CVAE, which utilize a latent space to capture observed sample distributions, have been applied to tasks like question generation \citep{CreativityDiverseQuestion} and image captioning \citep{wang2017diverseandaccurate} to foster diverse text generation. Models such as Wasserstein Autoencoders (WAE) \citep{gu2019dialogwae}, a variant of VAE, and Normalizing Flows \citep{dialogueNormFlow} are used in dialogue systems to generate a wider array of responses.

Another line of work has focused on the decoding phase of seq2seq models. \citeauthor{vijayakumar2018diverse} \citep{vijayakumar2018diverse} and \citeauthor{DistinctN-gram} \citep{DistinctN-gram} enhance the diversity of output by altering the objective function for decoding. Noisy Parallel Approximate Decoding (NPAD) \citep{cho2016noisy} injects random noise into the decoder's hidden state at each step. Alternative approaches, such as Top-g capping \citep{li2016simple} and Cluster Beam Search \citep{TAM2020101094}, set constraints or add selection strategies to beam search, encouraging the model to choose from more diverse candidates.

Other code-related domains, such as code generation from natural language \citep{wang2023codet5}, automatic program repair \citep{learnAPRSurvey}, and unit test generation \citep{diverseUnitTestGen}, have adopted the strategy of generating multiple candidate solutions for a single input and then selecting the best-performing candidate. Furthermore, some studies have demonstrated that enhancing diversity can significantly improve task success; for example, in unit test generation \citep{diverseUnitTestGen}, increased diversity has been shown to boost test suite coverage. However, to the best of our knowledge, no study has yet explored diverse generation in automatic source code summarization.

\section{Conclusion and Future Work}
In this paper, we introduced VPT, a novel approach that integrates CVAE into existing pre-trained models for code summarization without the need to retrain the entire model. The comprehensive experiment results demonstrated that our method can be easily adapted to a variety of pre-trained models, enhancing their ability to generate diverse sets of code summaries that are more likely to contain a more accurate summary. For reproducibility, the implementation of our approach is available in a repository\footnote{https://github.com/jundaz/VPT.git}.

For future work, we plan to explore integrating VPT with other LLMCs, including decoder-only models. The upper bound given by the set of 100 generated summaries also shown that we can benefit from even better subset selection or reranking methods to further exploit the potential of generated summaries. Additionally, conducting a user study would provide valuable insights into the practical applicability of our approach in real-world usage scenarios. We also aim to investigate alternative generative model architectures beyond CVAE and explore the combination of these with other parameter-efficient fine-tuning techniques. Finally, we plan to extend our method to additional code-related tasks, such as code generation and code translation, broadening the scope and applicability of our approach.




\bibliographystyle{elsarticle-num-names} 
\bibliography{VPT}






\end{document}